\documentclass[preprint,groupedaddress,showpacs,superscriptaddress,amssymb,amsmath]{revtex4-1}
\usepackage{graphicx,epsfig,epstopdf}
\usepackage{hyperref}
\usepackage{color}
\usepackage{threeparttable}
\usepackage{booktabs}
\usepackage[large]{subfigure}
\newcommand{\be}{\begin{equation}}
\newcommand{\ee}{\end{equation}}
\newcommand{\bea}{\begin{eqnarray}}
\newcommand{\eea}{\end{eqnarray}}

\color{black}

\begin{document}
\title{Stochastic Liouville Equations for  Femtosecond Stimulated Raman Spectroscopy}

\author{Bijay Kumar Agarwalla}
\affiliation{Department of Chemistry, University of California, Irvine, California 92617,
USA}
\author{Hideo Ando}
\affiliation{Department of Chemistry, University of California, Irvine, California 92617,
USA}
\author{Konstantin E. Dorfman}
\affiliation{Department of Chemistry, University of California, Irvine, California 92617,
USA}
\author{Shaul Mukamel}
\affiliation{Department of Chemistry, University of California, Irvine, California 92617,
USA}

\date{\today}

\begin{abstract}
Electron and vibrational dynamics of molecules are commonly studied by subjecting them to two interactions with a fast actinic pulse that prepares them in a nonstationary state and after a variable delay period $T$, probing them  with a Raman process  induced by  a combination of a broadband and a narrowband pulse. This  technique  known as  femtosecond stimulated Raman spectroscopy (FSRS) can effectively probe time resolved vibrational resonances. We show how FSRS signals  can be modeled and interpreted  using the stochastic Liouville equations (SLE) originally  developed for NMR lineshapes. The SLE provides a  convenient simulation protocol that  can describe complex dynamics due to coupling to collective coordinates  at much lower cost that a full dynamical simulation.
The origin of the dispersive features which appear when there is no  separation of timescales between vibrational variations and  dephasing is clarified. 

\end{abstract}

\maketitle
\section{Introduction}

Stimulated Raman spectroscopy is a common established time-resolved technique for monitoring vibrational motions \cite{Miz97,McCamant:JPCA:2003,Lee:JCP:2004,Fang:Nature:2009,Kukura:Science:2005,Kukura:AnnurevPhysChem:2007,Takeuchi:Science:2008,Kur12}. Multidimensional Raman techniques \cite{Zan09,biggs12,biggs13} probe the molecular system at multiple time points via a sequence of Raman processes, which measure correlations between several coherence periods. Non-adiabatic relaxation dynamics in chemical reactions  \cite{Mohammed:Science:2005,Schreier:Science:2007,Kukura:AnnurevPhysChem:2007}  as well as structural changes  
\cite{Kur12,Kukura:Science:2005,Takeuchi:Science:2008,Liebel:arXiv:2013} can then be probed with high temporal resolution. 
In a typical UV-(or visible) pump -  Raman probe  experiment, an actinic pump pulse launches a photochemical process in an excited electronic state,
 which is subsequently probed by a   delayed Raman pulse sequence.
Several variants of spontaneous and stimulated Raman probe-techniques which show high temporal and spectral resolution  have been reported \cite{McCamant:RevSci:2004,Kraack:PCCP:2013}. 
In the femtosecond stimulated Raman spectroscopy (FSRS) technique the Raman probe sequence consists of a picosecond pulse $\mathcal{E}_2$ superimposed with a femtosecond laser pulse $\mathcal{E}_3$ which stimulates the Raman signal. Starting from the earlier work of Yoshizawa and Kurosawa \cite{Yoshizawa: 1999} this technique has shown to be a sensitive local probe for ultrafast photo-induced processes \cite{Kukura:AnnurevPhysChem:2007, Fang:Nature:2009}. Different configurations of the FSRS techniques including temporally and spectrally overlapping pulses and resonant Raman processes \cite{Sun:2008}
and cascading effects in FSRS \cite{Zhao:2011} have been calculated \cite{Lee:JCP:2004}.

Typically in off-resonant FSRS a spectrally resolved pattern  of narrow 
vibrational lines (linewidth $\approx$10~cm$^{-1}$) is recorded in short time  intervals (20 fs).
FSRS is thus considered an ideal probe for ultrafast light-induced processes \cite{Kukura:AnnurevPhysChem:2007,Fang:Nature:2009}  which relates nuclear rearrangements to spectral changes.

Recently  we investigated the microscopic origin of the temporal and spectral resolution of FSRS \cite{Dorfman:PCCP:2013}. We proposed  three levels of theory which form a hierarchy of approximations for the simulation of the matter response   \cite{Dorfman:PCCP:2013} based on loop diagrams in the
frequency domain. These include sum over states and direct propagation of non adiabatic dynamics which is usually described by semiclassical models where bath degrees of freedom are treated classically.
The simplest level of modeling assumes that the vibrational frequency becomes time dependent due to e.g structural rearrangement \cite{Kukura:AnnurevPhysChem:2007,Dorfman:PCCP:2013}. In this case the  frequency trajectory is inserted into the wavefunction description of the signal and the bath degrees of freedom are not explicitly included in the Hilbert space description. A  higher level of theory assumes that some Hamiltonian parameters are fluctuating due to coupling to a bath described by collective coordinates. In that case the signals can be described using the stochastic Liouville equations (SLE) that act in the joint system plus bath space \cite{Kubo1, Kubo2,And54}. If the bath is harmonic and only modulates the frequencies one can solve the dynamics analytically using the cumulant expansion since the fluctuations are Gaussian, and avoid the SLE (see Eq. (39) of Ref. \cite{Dorfman:PCCP:2013}). However a  much broader class of models with continuous or discrete collective coordinates which may be coupled to any Hamiltonian parameter (not just the frequencies) can be treated  by SLE. Adding the bath complicates the calculation but it is still less demanding than the complete microscopic simulation involving all relevant degrees of freedom. The SLE, originally developed by Kubo and Anderson in NMR \cite{Kubo1, Kubo2,And54}, thus provide an affordable and practical level of modeling of complex lineshapes. In our earlier work we applied the SLE for comparison of the resolution of several Raman techniques for a simple two-state kinetics model \cite{Dorfman:JCP:2013}. Recently we have applied the SLE for identifying the spectroscopic signature of the underlying bond splitting mechanism in cyclobutane thymine dimer, one of the major lesions in DNA \cite{Ando:2014}. In the present work we provide more in depth general analysis of the SLE in the context of Stimulated Raman Spectroscopy and show how to use the SLE to trace the origin of the dispersive lineshapes reported in FSRS experiments.





\section{Transient Absorption of a shaped pulse; linear analogue of FSRS}

We first present a formal expression for the pump-probe signal defined as the change in the frequency-dispersed probe intensity .
We consider a multilevel quantum system described by the
Hamiltonian $H_0$ and coupled to an external optical field by the interaction  
\be
H_{\rm int} (t)= {\cal E}(t) V^{\dagger} +{\cal E}^*(t) V
\ee
where ${\cal E}(t)$ and ${\cal E}^*(t)$ are the positive and negative frequency components
of the total electric field operator $\tilde{E}(t)=\mathcal{E}(t)+\mathcal{E}^{*}(t)$ respectively. The dipole operator is $\tilde{V}=V+V^{\dagger}$ where $V^{\dagger}$ ($V$) is the raising (lowering)
operator responsible for excitation (de-excitation) between the molecular states. The total Hamiltonian is given by 
\be
H_{T}(t)=H_{0}+H_{\rm int}(t)
\ee

We use superoperator notation \cite{Shaulbook,Rahav_review,Harbola} that allows to derive compact
expressions for the signal. With each Hilbert
space operator
$A$, we associate two superoperators, denoted as $A_L$ (left) and $A_R$ (right) defined
through their action on Hilbert space operator $X$ as $A_L X \equiv A X,\, A_R X
\equiv X A$. We further define the linear combinations of these superoperators $A_{+}=(A_L
+ A_R)/2$
and $A_{-}= A_L - A_R$. $A_{+}$ ($A_-$) superoperator in Liouville space corresponds to an
anticommutation (commutation) operation in Hilbert space. Using this notation the
heterodyne detected signal (frequency dispersed transmission of ${\cal E}_p$ centered around $t=t_0$) can be recast as 
\be
S(\omega,t-t_0, \Gamma)= \frac{2}{\hbar} {\rm Im} \Big[ {\cal E}_{p}^*(\omega) \int_{-\infty}^{\infty} dt\, e^{i \omega
(t-t_0)} \big \langle {\cal T} V_L(t) \exp \big(-\frac{i}{\hbar} \int_{-\infty}^{t} d\tau_1
H_{{\rm int}-}(\tau_1)  \big) \big\rangle \Big],
\label{first-definition}
\ee
where $\Gamma$ represents the set of parameters of the incoming fields. The angular bracket $\langle \cdots \rangle$ 
represents the average with respect to the initially
prepared molecular density matrix. 
We define the interaction picture
superoperator as $
A_{\nu}(t)\equiv
\exp({i {H_0}_{-}(t-\tau_0)}) A_{\nu} \exp({-i {H_0}_{-}(t-\tau_0)}),$ $\nu=L,R$.
We also define the retarded Liouville space evolution operator as  
${\cal G}(t-\tau_0)\!=\!(-i/\hbar)\,\theta(t-\tau_0) \exp\big[\!-\!\frac{i}{\hbar} H_{0 -}(t-\tau_0)\big]$.
In the frequency domain the propagator is given as
${\cal G}(\omega)\equiv  \int_{-\infty}^{\infty} dt e^{i \omega(t-\tau_0)} {\cal G}(t-\tau_0)= \frac{1}{\hbar}(\omega I -
\frac{1}{\hbar} H_{{0}-}+ i \epsilon)^{-1}$  where $\epsilon$ is an infinitesimal positive number used to satisfy the causality condition.
Finally ${\cal T}$ is the
time-ordering superoperator that orders superoperators in increasing time argument from right to left i.e.,
\be 
{\cal T} A_{\nu} (t_1) B_{\nu'} (t_2)= \theta(t_1-t_2) A_{\nu}(t_1) B_{\nu'}(t_2) +
\theta(t_2-t_1) B_{\nu'}(t_2) A_{\nu}(t_1),  \quad \nu,\nu'=L,R.
\ee
Specific signals are obtained by expanding Eq.~(\ref{first-definition}) to the desired order in the field.

This paper focuses on the FSRS signal, which is a six-wave mixing process and thus we expand Eq.~(\ref{first-definition}) to fifth order in $H_{\rm int -}$. Before we proceed to FSRS however, we discuss a lower order signal which uses the same probe and may provide similar type of physical and chemical information. This is a Transient Absorption of the shaped pulse (TASP) with a
visible-pump and shaped probe (broad plus narrow band) which is a linear analogue of the FSRS and is obtained by expanding Eq.~(\ref{first-definition}) to third order in $H_{\rm int -}$. The  measurement is the transient absorption (TA) of a  shaped pulse consisting of a  narrow plus broadband probe (Fig.~1). 
An actinic pulse ${\cal E}_a$, centered at
$\tau_0$, promotes the molecule from the ground electronic state $|g\rangle$ to a superposition of
vibrational levels
of an electronic excited state. A spectrally narrow ${\cal E}_1$
and spectrally broad ${\cal E}_p$ probe pulse, both centered around $t_0>\tau_0$, then interact
with the molecule and can either stimulate emission to lower vibrational state or
absorption to a higher vibrational state and the frequency dispersed
transmission of the probe pulse is recorded. The positive frequency component of the
electric field is given by
\be
{\cal E}(t)={\cal E}_{a}(t-\tau_0)+ {\cal E}_1(t-t_0)+{\cal E}_p(t-t_0).
\ee
The entire sequence of events (including the actinic pulse) is a four wave mixing process and the signal scale as  ${\cal E}_1
{\cal E}_p {\cal E}_a^2$.  The signal can be viewed as a
generalized linear response with respect to ${\cal E}_1 {\cal E}_p$ from the non-stationary
state prepared by ${\cal E}_a$.

\begin{figure}
\includegraphics[width=0.5\columnwidth]{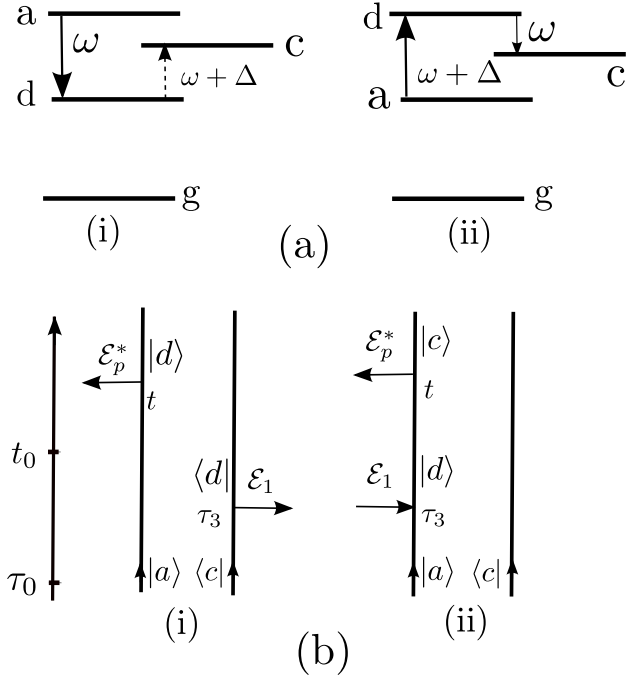}
\caption{TASP signal: Level scheme (a) and ladder diagrams \big(Eq.~(\ref{TA-1}) and Eq.~(\ref{TA-first})\big) (b).} 
\end{figure}
Fig.~1  represents the level scheme and corresponding ladder diagrams that contribute to the signal (for
diagram rules see ref.~\cite{Harbola}). We first expand the exponential in Eq.~(\ref{first-definition}) to first order in ${\cal
E}_1$ and ${\cal E}_p$ and to second order in ${\cal E}_a$, taking the actinic pulses to be impulsive ${\cal E}_a(\tau)= {\cal E}_a\delta(\tau)$. The following expressions can be directly read off the diagrams (see Appendix A)
\bea
S_{TASP}^{(i)}(\omega;t_0,\tau_0)&=& \frac{2}{\hbar} \,{\rm Im} \Big[
(-\frac{i}{\hbar})\int_{-\infty}^{\infty}
dt \int_{0}^{\infty} dt_3 \,e^{i \omega (t-t_0)}
\,\langle V_L {\cal G}(t_3) V_{R}^{\dagger} {\cal G}(t\!-\!t_3\!-\!\tau_0) \rangle^{'} \nonumber \\
&&\times \,{\cal E}_{p}^{*}(\omega) {\cal E}_{1}(t\!-\!t_3\!-\!t_0)\Big],
\label{TA-1}
\eea
\bea
S_{TASP}^{(ii)}(\omega;t_0,\tau_0)&=& \frac{2}{\hbar} \,{\rm Im} \Big[
(\frac{i}{\hbar})\int_{-\infty}^{\infty}
dt \int_{0}^{\infty} dt_3 \,e^{i \omega (t-t_0)}
\,\langle V_L {\cal G}(t_3) V_{L}^{\dagger} {\cal G}(t\!-\!t_3\!-\!\tau_0) \rangle^{'}  \nonumber \\
&&\times \,{\cal E}_{p}^{*}(\omega) {\cal E}_{1}(t\!-\!t_3\!-\!t_0)\Big].
\label{TA-first}
\eea
where $\langle \cdots \rangle' \equiv \sum_{ac} \rho_{ac} \langle \langle I| \cdots |ac\rangle \rangle$.  If the system is prepared in a superposition of vibrational states $a$ and $c$ in an electronic excited state, created by two interactions with actinic pulse $\mathcal{E}_a$ represented by the action of $V_R V_L^{\dagger}$ on the ground electronic
state. A non-stationary vibrational wave-packet in the excited state \cite{Lee:JCP:2004}  is then described by $|\mathcal{E}_a|^2V_R V_L^{\dagger}
|gg\rangle \rangle= \sum_{ac} \rho_{ac} |a c\rangle \rangle$ where $\rho_{ac}= |{\cal E}_a|^{2} \mu_{ga} \mu_{cg}$. By
 inverse Fourier transformation of the propagator and the electric field, performing time integrals and introducing a delay between preparation time $\tau_0$ and probe pulse $t_0$: $T=t_0-\tau_0$ Eqs. (\ref{TA-1}) - (\ref{TA-first})  yield
\be
S_{TASP}^{(i)}(\omega;T)= \frac{2}{\hbar} \, {\rm Im} \Big[(-\frac{i}{\hbar})
\int_{-\infty}^{\infty}
\frac{d\Delta}{2\pi} \,
e^{-i\Delta T} {\cal
E}_{p}^{*}(\omega) {\cal E}_{1}(\omega+\Delta) 
\tilde{\chi}_{TASP(i)}^{(1)}(-\omega,\omega+\Delta)\Big],
\label{final-linear-signal-1}
\ee
\be
S_{TASP}^{(ii)}(\omega;T)= \frac{2}{\hbar} \, {\rm Im} \Big[(\frac{i}{\hbar})
\int_{-\infty}^{\infty}
\frac{d\Delta}{2\pi} \,
e^{-i\Delta T} {\cal
E}_{p}^{*}(\omega) {\cal E}_{1}(\omega+\Delta) 
\tilde{\chi}_{TASP(ii)}^{(1)}(-\omega,\omega+\Delta)\Big],
\label{final-linear-signal-2}
\ee
where we have introduced the generalized susceptibility $\tilde{\chi}_{TASP(j)}^{(1)}(-\omega,\omega+\Delta)$, $j=i,ii$
\be
\tilde{\chi}_{TASP(i)}^{(1)}(-\omega,\omega+\Delta) = 
\langle V_L {\cal
G}(\omega) V_{R}^{\dagger}{\cal
G}(-\Delta)\rangle^{'},
\label{chia1}
\ee
\be
\tilde{\chi}_{TASP(ii)}^{(1)}(-\omega,\omega+\Delta) = 
\langle V_L {\cal
G}(\omega) V_{L}^{\dagger}{\cal
G}(-\Delta)\rangle^{'}.
\label{chia3}
\ee
Four-wave-mixing such as transient absorption was proposed by P. Champion's group \cite{Kumar:2001,Kumar2:2001} where it has been called ``effective linear response approach''. Here we extend the idea to recasting the $n+m$-wave mixing signal in terms of the effective $m$-wave mixing.
We use this approach in the context of six-wave mixing FSRS signal. Eqs. (\ref{final-linear-signal-1}) - (\ref{final-linear-signal-2}) are similar to those obtained by Champion, but for TASP - the linear analogue of the FSRS.

The signal may  be interpreted as a generalized linear response to 
the field ${\cal E}_1 {\cal E}_p$ from a nonstationary state $\rho_{ac}$ prepared by the actinic pulse
${\cal E}_a$. Note that $\tilde{\chi}_{TASP}^{(1)}$ depends on two frequencies (rather than one for systems initially at equilibrium). This 
implies that
the signal (Eqs.~(\ref{final-linear-signal-1}, \ref{final-linear-signal-2})) is sensitive to the phase of the
field. The choice of broadband $\mathcal{E}_p$ and narrowband $\mathcal{E}_1$ is one  example of more broadly defined pulse shaping \cite{Dantus1, Dantus2, Dantus3, Dantus4}. Different phase shapes can manipulae signals by enhancing or suppressing various spectral features and changing the line shapes, etc. In our earlier work we have been studying the pulse shaping of linear signals as a possible tool for coherent control \cite{Shaul_control} . Our formalism can incorporate arbitrary pulse shapes which can and provide many novel control tools for the signals.

Assuming that the ${\cal E}_1$ pulse is spectrally narrow (picosecond) i.e., ${\cal
E}_1(\omega+\Delta)={\cal E}_1 \delta(\omega+\Delta-\omega_1)$ we can perform the integral over
$\omega_1'$ which results in
  \be
S^{(i)}_{TASP}(\omega;T)= \frac{2}{\hbar} \,{\rm Im} \Big[(-\frac{i}{\hbar})  \,
e^{i(\omega-\omega_1)T} {\cal
E}_{p}^{*}(\omega) {\cal E}_{1} 
\tilde{\chi}_{TASP(i)}^{(1)}(-\omega,\omega_1)\Big],
\ee
\be
S^{(ii)}_{TASP}(\omega;T)= \frac{2}{\hbar} \,{\rm Im} \Big[(\frac{i}{\hbar})  \,
e^{i(\omega-\omega_1)T} {\cal
E}_{p}^{*}(\omega) {\cal E}_{1} 
\tilde{\chi}_{TASP(ii)}^{(1)}(-\omega,\omega_1)\Big].
\ee
The total signal is finally given by $S_{TASP}^{(\rm total)}(\omega;T)=S^{(i)}_{TASP}(\omega;T)+S^{(ii)}_{TASP}(\omega;T)$.
In the absense of a bath the correlation functions can be expanded in terms of molecular eigenstates which gives the signal as 
%
\bea
S_{TASP}^{(i)}(\omega;T)&=&\frac{2}{\hbar} \, {\rm Im}\sum_{a,c,d} \Bigg
[(-\frac{i}{\hbar}) 
\frac{e^{-i (\omega-\omega_1)T} \mu_{da} \mu_{cd} {\cal E}_p^{*}(\omega) {\cal E}_1 \rho_{ac}} {(\omega-\omega_{ad}+i\gamma_{ad})(\omega-\omega_1-\omega_{ac}+i\gamma_{ac})}
\Bigg], \nonumber \\
S_{TASP}^{(ii)}(\omega;T)&=&\frac{2}{\hbar} \,{\rm Im}\sum_{a,c,d} \Bigg
[(\frac{i}{\hbar}) 
\frac{e^{-i (\omega-\omega_1)T} \mu_{ad} \mu_{dc} {\cal E}_p^{*}(\omega) {\cal E}_1
\rho_{ac}}{(\omega-\omega_{da}+i\gamma_{da})(\omega-\omega_1-\omega_{ca}+i\gamma_{ca})}
\Bigg].
\label{eq:STA}
\eea
These results will later be compared with the FSRS signal Eq.~(\ref{eq:FSRS2}).
\section{off-resonant FSRS}

\begin{figure}
\includegraphics[width=0.35\columnwidth]{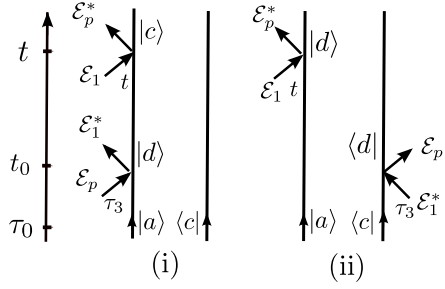}
\caption{Ladder diagrams for FSRS signal Eq.~(\ref{FSRS-1}) and Eq.~(\ref{FSRS-first}).}
\end{figure}
FSRS is a powerful technique
for studying photophysical and photochemical processes in molecules \cite{Kukura:AnnurevPhysChem:2007, photo2} including  vibrational and nonadiabatic electron dynamics . 
It is a six-wave mixing process \cite{Dorfman:JCP:2013} that, when the actinic action is treated implicitly, is given by a generalized $\tilde{\chi}^{(3)}$. Moreover in the electronically off resonant regime, considered here, it can
be expressed in terms of a generalized linear response $\tilde{\chi}^{(1)}$ that resembles TASP. 

The general expression for a four wave-mixing probe of a nonstationary system is given in Appendix B. FSRS (Fig.~2) can be described using
diagrams similar  to Fig.~(1) where we simply replace the dipole operators by the
polarizibility. We use the same level scheme as in Fig.~(1). The
effective radiation-matter interaction Hamiltonian  (in the Rotating wave
approximation) for the electronically
off-resonant Raman process induced by pump pulse ${\cal E}_1$ and the probe pulse ${\cal E}_p$,  reads
\be
H_{\rm int}(t)= \alpha_{n} {\cal E}_{p}^{*}(t-t_0)\,{\cal E}_1(t-t_0) + {\cal
E}_{a}(t-\tau_0)\, V_e^{\dagger} + h.c.
\ee
Here the dipole operator $V_{e}$ couples the molecule with actinic pulse ${\cal
E}_a$, centered at time $\tau_0$ and creates a non-stationary state in the excited
electronic level. $\alpha_n=
\tilde{\alpha}_n+
\tilde{\alpha}_n^{\dagger}$
is the excited state polarizibility that couples parametrically the pump  ${\cal E}_1$
and the probe fields ${\cal E}_p$. Both
pulses arrive simultaneously and are centered at time $t_0 > \tau_0$. Similar to TASP we set the delay between the actinic and Raman
pulses as $T=t_0-\tau_0$. Note that this off-resonant Raman process is instantaneous and the system can only spend a very short time in the intermediate electronic
state. Following
Eq.~(\ref{first-definition}), the frequency-dispersed probe transmission signal
is written as 
\be
S_{FSRS}(\omega;T)= \frac{2}{\hbar} \, {\rm Im} \Big[ \int_{-\infty}^{\infty} dt
\int_{-\infty}^{\infty}
\frac{d\omega_1'}{2\pi}
e^{i (\omega-\omega_1')
(t-t_0)} {\cal E}_{p}^*(\omega) {\cal E}_1(\omega_1') \langle \alpha_L(t)e^{-(i/\hbar)\int H_{int-}(\tau)d\tau}
 \rangle \Big].
\label{second-definition}
\ee
The picosecond pump
pulse ${\cal E}_1$ has a narrow bandwidth centered at frequency
$\omega_1$. This signal can be read off from the diagrams given in Fig.~(2) and is given as
\bea
S_{FSRS}^{(i)}(\omega;t_0,\tau_0)&=& \frac{2}{\hbar} \,{\rm Im} \Big[
(\frac{i}{\hbar})\int_{-\infty}^{\infty}
dt \int_{0}^{\infty} dt_3 \,e^{i \omega (t-t_0)}
\,\langle \alpha_{nL} {\cal G}(t_3) \alpha_{nL} {\cal G}(t\!-\!t_3\!-\!\tau_0) \rangle^{'} \nonumber \\
&&\times \,{\cal E}_{p}^{*}(\omega)\mathcal{E}_p(t-t_3-t_0) \mathcal{E}_1(t){\cal E}_{1}^{*}(t\!-\!t_3\!-\!t_0)\Big],
\label{FSRS-1}
\eea
\bea
S_{FSRS}^{(ii)}(\omega;t_0,\tau_0)&=& \frac{2}{\hbar} \,{\rm Im} \Big[
(-\frac{i}{\hbar})\int_{-\infty}^{\infty}
dt \int_{0}^{\infty} dt_3 \,e^{i \omega (t-t_0)}
\,\langle \alpha_{nL} {\cal G}(t_3) \alpha_{nR} {\cal G}(t\!-\!t_3\!-\!\tau_0) \rangle^{'} \nonumber \\
&&\times \,{\cal E}_{p}^{*}(\omega)\mathcal{E}_p(t-t_3-t_0)\mathcal{E}_1(t) {\cal E}_{1}^{*}(t\!-\!t_3\!-\!t_0)\Big].
\label{FSRS-first}
\eea
%
Eqs. (\ref{FSRS-1})- (\ref{FSRS-first}) for FSRS or Eqs.(\ref{TA-1})-(\ref{TA-first}) for TASP can be used for high level numerical simulations including nonadiabatic dynamics. Alternatively we can recast the signal using frequency domain Green's functions. Denoting $T=t_0-\tau_0$ we get the total signal as $S_{FSRS}^{\rm total}(\omega;T)=S_{FSRS}^{(i)}(\omega;T)+S_{FSRS}^{(ii)}(\omega;T)$,
\be
S_{FSRS}^{(i)}(\omega;T)= \frac{2}{\hbar} \,{\rm Im} \Big[(\frac{i}{\hbar})  \int_{-\infty}^{\infty}
\frac{d\Delta}{2\pi} e^{-i\Delta T}
|{\cal E}_{1}|^2 {\cal
E}_p^{*}(\omega) {\cal E}_{p}(\omega+\Delta)  \tilde{\chi}_{FSRS(i)}^{(1)}(-\omega\!+\!\omega_1,-\omega_1\!+\!\omega\!+\!\Delta)
\Big],
\label{eq:FSRS01}
\ee
\be
S_{FSRS}^{(ii)}(\omega;T)= \frac{2}{\hbar} \,{\rm Im} \Big[ (-\frac{i}{\hbar})\int_{-\infty}^{\infty}
\frac{d\Delta}{2\pi} e^{-i\Delta T}
|{\cal E}_{1}|^2 {\cal E}_p^{*}(\omega)
{\cal E}_{p}(\omega+\Delta)  \tilde{\chi}_{FSRS(ii)}^{(1)}(-\omega\!+\!\omega_1,-\omega_1\!+\!\omega\!+\!\Delta) \Big],
\label{eq:FSRS02}
\ee
The generalized  FSRS susceptibility is now given by  
\be
\tilde{\chi}_{FSRS(i)}^{(1)}(-\omega\!+\!\omega_1,-\omega_1\!+\!\omega\!+\!\Delta) = \big
\langle \alpha_{nL} {\cal G}(\omega-\omega_1) \alpha_{nL} {\cal
G}(-\Delta) \big \rangle^{'},
\ee
\be
\tilde{\chi}_{FSRS(ii)}^{(1)}(-\omega\!+\!\omega_1,-\omega_1\!+\!\omega\!+\!\Delta) = \big \langle \alpha_{nL}
{\cal G}(\omega-\omega_1) \alpha_{nR} {\cal
G}(-\Delta) \big \rangle^{'}.
\ee
This signal may be viewed as the linear response to ${\cal E}_p^*\, {\cal E}_1$ with the
generalized susceptibility formally similar to TASP (Eqs.(\ref{chia1}) - (\ref{chia3})) except that the dipole operators $V$ are
replaced by the polarizability $\alpha_n$. The reduced description of the six-wave mixing FSRS in terms of an effective four-wave-mixing has been studied previously \cite{Marek} where the response function has been calculated using the multimode Brownian oscillator model. The Green's functions expressions presented here provide more general framework for calculating the signals. In particular our approach is very effective if the excited state coherence is long lived on the time scale of the Raman probe sequence \cite{Kraack, Takeuchi:Science:2008}. It also applies in the presence of a non- trivial relaxation process such as internal conversion between excited states \cite{Marek}.
In the absense of a bath we expand the signals (\ref{eq:FSRS01}) - (\ref{eq:FSRS02}) in the molecular eigenstate basis and  obtain
\bea
S_{FSRS}^{(i)}(\omega;T)&=&\frac{2}{\hbar} \, {\rm Im}\sum_{a,c,d} \Bigg
[(\frac{i}{\hbar}) \int_{-\infty}^{\infty}\frac{d\Delta}{2\pi}
\frac{e^{-i\Delta T} \alpha_{da} \alpha_{a'd} {\cal E}_p^{*}(\omega)\mathcal{E}_p(\omega+\Delta) |{\cal E}_1|^2
\rho_{ac}}{(\omega-\omega_1-\omega_{ad}+i\gamma_{ad})(-\Delta-\omega_{ac}+i\gamma_{ac})}
\Bigg], \nonumber \\
S_{FSRS}^{(ii)}(\omega;T)&=&\frac{2}{\hbar} \,{\rm Im}\sum_{a,c,d} \Bigg
[(-\frac{i}{\hbar})  \int_{-\infty}^{\infty}\frac{d\Delta}{2\pi}
\frac{e^{-i \Delta T} \alpha_{ad} \alpha_{dc} 
{\cal E}_p^{*}(\omega)\mathcal{E}_p(\omega+\Delta) |{\cal E}_1|^2
\rho_{ac}}{(\omega-\omega_1-\omega_{da}+i\gamma_{da})(-\Delta-\omega_{ca}+i\gamma_{ca})}
\Bigg].
\label{eq:FSRS2}
\eea
Comparing FSRS signal (Eq. (\ref{eq:FSRS2})) to the TASP (Eq. (\ref{eq:STA})), we note several differences. First the dipole moments in TASP are replaced by corresponding excited state polarizabilities in the case of FSRS. When the narrowband pulse $\mathcal{E}_1$ is monochromatic the TASP signal has a perfect frequency resolution, which governs the Raman resonance $\omega-\omega_1\sim\omega_{da}$. In the same time it loses all temporal resolution, since the exponent $e^{-i(\omega-\omega_1)T}$ in Eq. (\ref{eq:STA}) has no dependence on material parameters such as energies and dephasing rates. In order to obtain some temporal resolution, $\mathcal{E}_1$ must have a finite bandwidth. Generally  the temporal and spectral resolutions of the TASP signal will be Fourier conjugates, so the perfect time resolution would correspond to poor spectral resolution. In contrast, the FSRS signal contains an extra integration over $\Delta$. Therefore, the $ac$ state in Eq. (\ref{eq:FSRS2}) does not give a spectral signature but rather controls the temporal resolution. Furthermore, once we fix the narrowband $\mathcal{E}_1$ with frequency $\omega_1$ the Raman resonance $\omega-\omega_1\sim\omega_{da}$ is now well resolved. Thus under the same conditions FSRS has both high spectral and temporal resolution, which are not directly Fourier conjugates of each other. In the next section we show how the coupling to a dynamical bath can be incorporated to calculate signal using the stochastic Liouville equation.



\section{Simulating FSRS signal by the Stochastic Liouville Equations}
The stochastic Liouville equations (SLE), first developed for NMR, is a convenient tool for computing spectral line shapes \cite{Kubo1, Kubo2, Gamliel, Tanimura,Sanda}.  This approach assumes that the quantum system of interest is affected by a classical bath, whose stochastic dynamics is described by a Markovian master equation. The SLE is an equation of motion for the joint system plus bath density matrix $\rho$
\begin{equation}
\label{eq:sle}
\frac{d\rho} {dt}
=
\hat{\mathcal{L}} \rho(t)
=
-\frac{i} {\hbar} \left[ H, \rho(t) \right] + \hat{L} \rho(t).
\end{equation}
where $\hat{L}$ represents the stochastic Markovian dynamics of the bath. Both discrete $N$-state jump and continuous collective coordinates (Fokker Planck equations) are commonly used to model the bath. In our model, the system has two vibrational states $a$ and $c$, and the vibrational frequency $\omega_{ca,s}$ is perturbed by the bath which has $N$ states, $s=1,2,\cdots N$.  The entire density matrix $\rho$ thus has $4N$ components $\left| \nu \nu' s \right\rangle\rangle$ which represent the direct product of four Liouville space states $\left| \nu \nu' \right\rangle\rangle$, where $\nu,\nu'=a,c$, and $s$ represents $N$ bath states.  The Liouville operator $\hat{\mathcal{L}}$ is diagonal in the vibrational space, and is thus represented by four $N \times N$ diagonal blocks in bath space.
\begin{equation}
\label{eq:liouville}
[ \hat{\mathcal{L}} ]_{\nu \nu' s,\nu_{1} \nu'_{1} s'}
=
\delta_{\nu\nu_{1}}\delta_{\nu'\nu'_{1}} \left([\hat{L}_S]_{s,s'} + \delta_{ss'} [\hat{\mathcal{L}}_{S}]_{\nu \nu' s,\nu \nu' s} \right),
\end{equation}
where $\hat{L}_{S}=-K$ describes the kinetics given by the rate equation:
\begin{equation}
\label{eq:rateeq}
\frac{d} {dt} \rho_{aa}^{(s)}(t)
=
-\sum_{s'}K_{ss'} \rho_{aa}^{(s')}(t),
\end{equation}
where $\rho_{aa}^{(s)}(t)$ is the population of the $s$-th bath state. The solution of Eq. (\ref{eq:rateeq}) is given by
\begin{equation}
\label{eq:concentrations}
\rho_{aa}^{(s)}(t)
=
\sum_{s'} U_{ss'}
\exp{\left[-K^{\rm diag} t\right]_{s's'}}
U_{s's}^{-1}
\rho_{aa}^{(s)}(0),
\end{equation}
where $U$ is the transformation matrix, where the eigenvectors are organized as rows. This matrix satisfies left-eigen equation $\sum_{p}U_{sp}K_{ps'}=K^{\rm diag}_{s's'}U_{ss'}$ as the rate matrix $K$ is not Hermitian.  $\rho_{aa}^{(s)}(0)$ represents the population of the initial bath state.

The coherent part $\hat{\mathcal{L}}_{S}=-(i/\hbar)\left[ H_{S}, \ldots \right]$, which describes the vibrational dynamics, vanishes for the $\left| a a \right\rangle\rangle$ and $\left| c c \right\rangle\rangle$ blocks, $[\hat{\mathcal{L}}_{S}]_{aa,aa}=[\hat{\mathcal{L}}_{S}]_{cc,cc}=0$.  The remaining blocks of $\hat{\mathcal{L}}_{S}$ read
\begin{equation}
\label{eq:ls2}
[\hat{\mathcal{L}}_{S}]_{ac,ac}
=
i
\left(
    \begin{array}{cccc}
      \omega_{ca}^{(1)} & 0 & ... & 0 \\
      0 & \omega_{ca}^{(2)} & ... & 0 \\
      ...& ...& ...&...\\
      0 & 0 & ... & \omega_{ca}^{(N)} \\
    \end{array}
\right).
\end{equation}
{\small The two Liouville space Green's functions (i.e., the solution of Eq. (\ref{eq:sle})) are thus given by}
\begin{equation}
\label{eq:green1}
\begin{split}
\mathcal{G}_{aa,aa}(t)
&=
-\frac{i} {\hbar} \theta(t) \exp{\left[ [\hat{L}_{S}] t \right]}\\
&=
-\frac{i} {\hbar} \theta(t) \bm{U} \exp{\left[ [\hat{L}_{S}]^{\rm diag} t \right]} \bm{U}^{-1},
\end{split}
\end{equation}

\begin{equation}
\label{eq:green2}
\begin{split}
\mathcal{G}_{ac,ac}(t)
&=
-\frac{i} {\hbar} \theta(t) \exp{\left[ ([\hat{L}_{S}] + [\hat{\mathcal{L}}_{S}]_{ac,ac}) t \right]}\\
&=
-\frac{i} {\hbar} \theta(t) \bm{V} \exp{\left[ [\hat{\mathcal{L}}]_{ac,ac}^{\rm diag} t \right]} \bm{V}^{-1},
\end{split}
\end{equation}
{\small where $\bm{U}$ and $\bm{V}$ are transformation matrices, which diagonalize the matrices in the exponents.}

\begin{figure}[htbp]
\begin{center}
\subfigure[]{\includegraphics[scale=0.40,trim=0 0 0 70,angle=90]{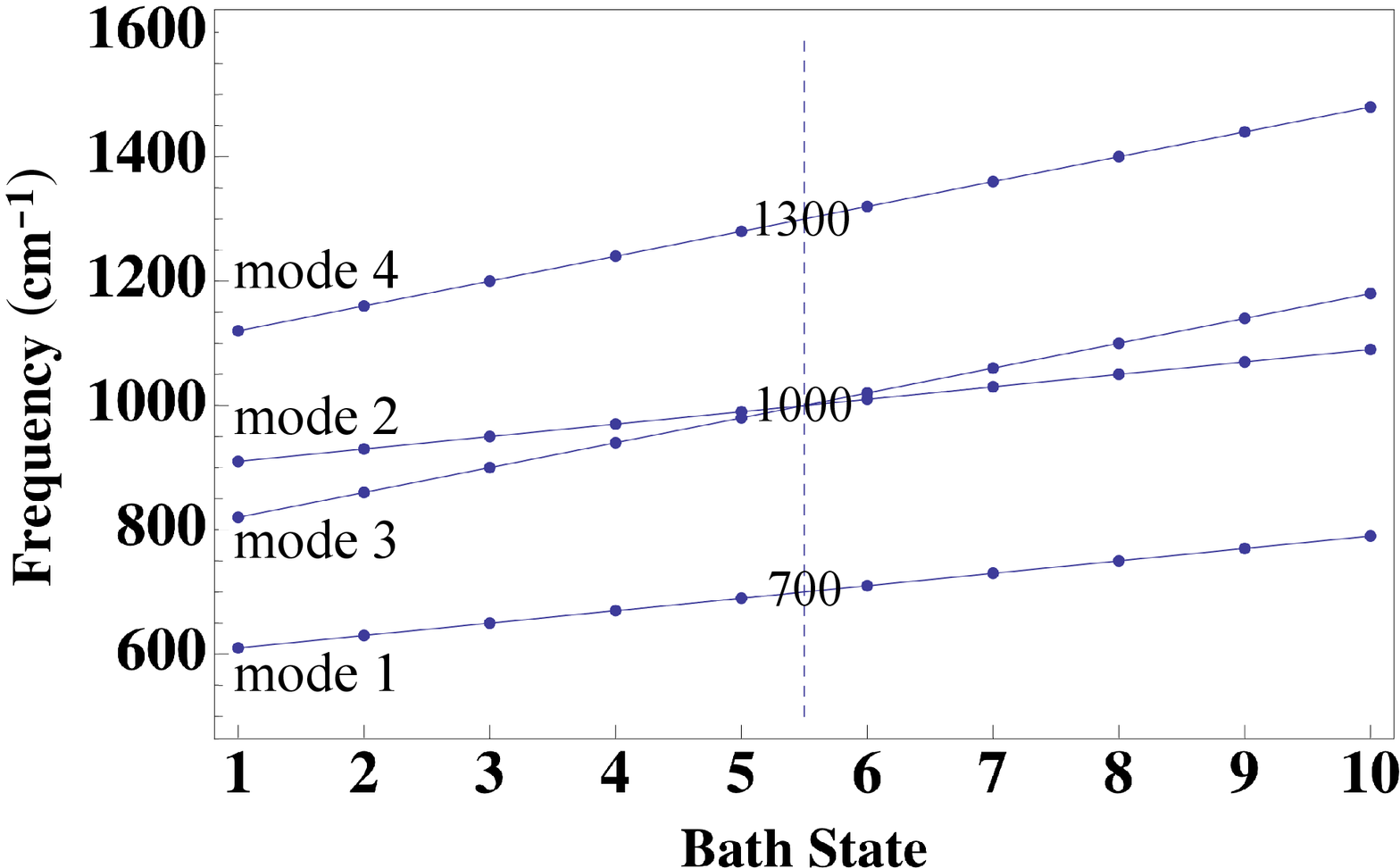}}
\vspace{20mm}
\subfigure[]{\includegraphics[scale=0.40,trim=0 0 0 -50,angle=90]{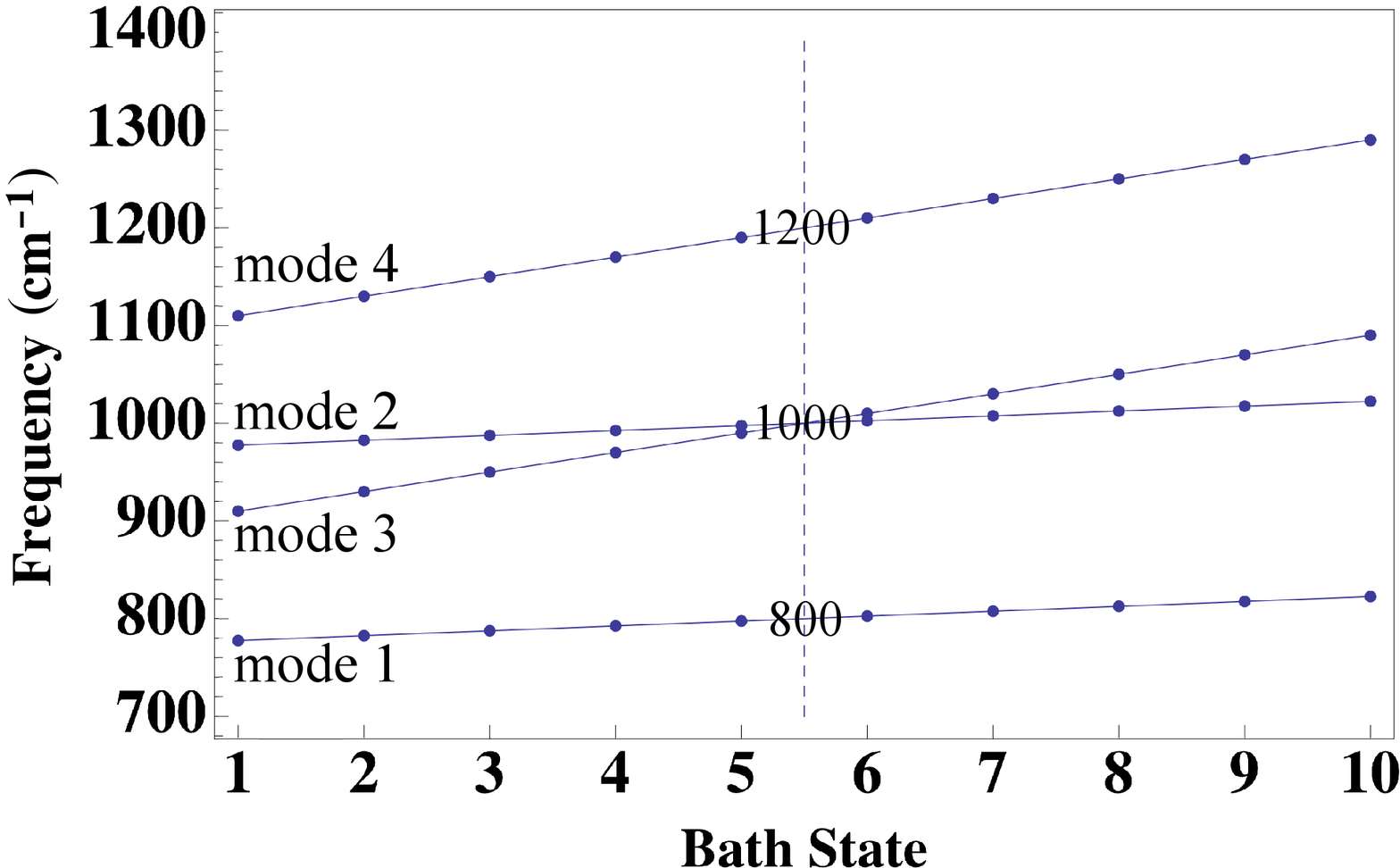}}
\vspace{-22mm}
\caption{Model frequency change of four vibrational modes along the sequential reaction.  (a) Parameter regime I and (b) regime II.  In Eq. (\ref{eq:omegachange}), the $\omega_{ca}^{(1)}$ of mode 2 and 3 are set so that their $\omega_{ca}^{(s)}$ frequencies show crossing during the transition from bath state 5 to 6.  The frequencies at the crossing points (dashed lines) are shown.}
\label{frequencychange}
\end{center}
\end{figure}

{\small The time domain FSRS signal on the Stokes side ($\omega<\omega_0$) is given by}
\begin{equation}
\label{eq:tdfsrs}
S_{FSRS}(\omega,T) = 
\Im \int_{-\infty}^{\infty} \frac{d\Delta} {2 \pi} \mathcal{E}_{p}^{*}(\omega) \mathcal{E}_{p}(\omega + \Delta)
\tilde{S}^{(i)}_{FSRS}(\omega,T;\Delta),
\end{equation}
{\small where $\tilde{S}^{(i)}_{FSRS}(\omega,T;\Delta)$ can be recast in Liouville space as follows:}
\begin{equation}
\label{eq:tdfsrstilde}
\begin{split}
\tilde{S}^{(i)}_{FSRS}(\omega,T;\Delta) = &
\frac{2} {\hbar}
\int_{-\infty}^{\infty} {dt} \int_{-\infty}^{t} {d\tau_{3}} |\mathcal{E}_{1}|^{2} |\mathcal{E}_{a}|^{2}
e^{-i \Delta (\tau_{3} - T)}\\
& \times e^{i (\omega - \omega_{1}) (t - \tau_{3})} \mathcal{F}(t - \tau_{3},\tau_{3}),
\end{split}
\end{equation}
{\small where by using the Green's functions in Eqs. (\ref{eq:green1}) and (\ref{eq:green2}), the matter correlation function $\mathcal{F}(t_{1},t_{2})$ is given by}
\begin{equation}
\label{eq:fmatter}
\begin{split}
\mathcal{F}(t_{1},t_{2}) =& 
-\frac{i} {\hbar}
\sum^{}_{a,c} \alpha^{2}_{ac} |V_{ag}|^{2} \left\langle\langle I \right| 
\mathcal{G}_{ac,ac}(t_{1})
\mathcal{G}_{aa,aa}(t_{2})
\left| \rho_{0} \right\rangle\rangle_{S}\\
=&
-( \frac{i} {\hbar})^{3}
\sum^{}_{a,c} \alpha^{2}_{ac} |V_{ag}|^{2} \theta(t_{1}) \theta(t_{2}) e^{-\gamma_{a} (t_{1} + 2 t_{2})}\\
& \times (1,1,...,1) \bm{V} \exp{\left[ [\hat{\mathcal{L}}]_{ac,ac}^{\rm diag} t_{1} \right]} \bm{V}^{-1} \bm{U} \exp{\left[ [\hat{L}_{S}]^{\rm diag} t_{2} \right]} \bm{U}^{-1} 
\left(
    \begin{array}{c}
      1 \\
      0 \\
      ... \\
      0
    \end{array}
  \right).
\end{split}
\end{equation}
{\small Here, the initial state is the direct product,}
\begin{equation}\label{eq:rho0}
\left| \rho_{0} \right\rangle\rangle_{S} = \left| a a \right\rangle\rangle \left(
    \begin{array}{c}
      1 \\
      0 \\
      ... \\
      0
    \end{array}
  \right),
\end{equation}
{\small and we have traced over the final state $\left\langle\langle I \right| = (1,1,...,1){\rm Tr}$ where ${\rm Tr}=\left\langle\langle a a \right| + \left\langle\langle c c \right|$.  Vibrational dephasing terms have been added; $e^{-\gamma_{a} t}$ is added to $\mathcal{G}_{ac,ac}$ and $e^{-2 \gamma_{a} t}$ to $\mathcal{G}_{aa,aa}$. Inhomogeneous broadening can be included by convoluting the present results with a spectral distribution. Gaussian frequency fluctuations can be incorporated via the cumulant expansion which solves the SLE for a Brownian oscillator bath. We have incorporated this level of theory in an earlier study \cite{Dorfman:PCCP:2013}.

Upon evaluating the time integrals in Eq. (\ref{eq:tdfsrstilde}) we obtain
\begin{equation}
\label{eq:tdfsrstilde2}
\begin{split}
&\tilde{S}^{(i)}_{FSRS}(\omega,T;\Delta) =\\
&\frac{-2i}{\hbar^2}|\mathcal{E}_1|^2|\mathcal{E}_a|^2\sum_{a,c}\alpha_{ac}^2 |V_{ag}|^2 e^{i\Delta T}\langle\langle I|\mathcal{G}_{ac,ac}(\omega-\omega_1)\mathcal{G}_{aa,aa}(-\Delta)|\rho_0\rangle\rangle_S.
\end{split}
\end{equation}
Following Eq. (\ref{eq:concentrations}) we introduce the bath population of the state $a$ after interaction with the actinic pulse:
\begin{equation}\label{eq:rhoaa}
\rho_{aa}^{(s)}(t)=|\mathcal{E}_a|^2 |V_{ag}|^2 \mathcal{G}_{aa,aa}(t)|\rho_0\rangle\rangle_S.
\end{equation}
Substituting Eq. (\ref{eq:rhoaa}) into the signal expression Eq.~(\ref{eq:tdfsrstilde}) and Eq.~(\ref{eq:tdfsrs}) gives
 \begin{align}\label{eq:FSRSi0}
S_{FSRS}(\omega,T)&=\Im\frac{-2i}{\hbar^2}\mathcal{E}_p^{*}(\omega)|\mathcal{E}_1|^2\sum_{a,c}\alpha_{ac}^2\sum_{s}\mathcal{G}_{ac,ac,s}(\omega-\omega_1)\int_{-\infty}^{\infty}\frac{d\Delta}{2\pi}\mathcal{E}_p(\omega+\Delta)e^{i\Delta T}\rho_{aa}^{(s)}(-\Delta),
\end{align}
where $\rho_{aa}(-\Delta)$ is the Fourier transform of the population of the state $a$. $\mathcal{G}_{ac,ac}(\omega)$ is a frequency-domain Green's function and $\sum_s$ represents the sum over bath states. It follows from Eq. (\ref{eq:FSRSi0}) that the $\Delta$ integration represents a path integral over the bandwidth corresponding to the inverse dephasing time scale. This integral is generally a complex number. Therefore, the signal (\ref{eq:FSRSi0}) depends on both real and imaginary parts of the coherence Green's function $\mathcal{G}_{ac,ac}(\omega)$, and thus contains absorptive as well as dispersive spectral features. In the limit of slow fluctuations, one can neglect the jump dynamics during the dephasing time. In this case we can replace $\mathcal{E}_p(\omega+\Delta)\simeq\mathcal{E}_p(\omega)$, the integral over $\Delta$ simply yields $\rho_{aa}(T)$ and we obtain the static averaged signal
 \begin{align}\label{eq:staticlimit}
S_{FSRS}(\omega,T)&=\sum_{a}\sum_sS_{FSRS,a}^{(s)}(\omega)\rho_{aa}^{(s)}(T),
\end{align}
with
\begin{align}
S_{FSRS,a}^{(s)}(\omega)=-\Re\frac{2}{\hbar^2}|\mathcal{E}_p(\omega)|^2|\mathcal{E}_1|^2\sum_{c} \alpha_{ac}^{2}\mathcal{G}_{ac,ac,s}(\omega-\omega_1).
\end{align}
For comparison we give the corresponding TASP signal:
 \begin{align}\label{eq:FSRSi0TA}
S_{TASP}(\omega,T)&=\Im\frac{-2i}{\hbar^2}\mathcal{E}_p^{*}(\omega)\mathcal{E}_1\sum_{a,c}|\mu_{ac}|^2\sum_{s}\mathcal{G}_{ac,ac,s}(\omega)e^{i(\omega-\omega_1)T}\rho_{aa}^{(s)}(\omega-\omega_1).
\end{align}
Unlike the general FSRS signal (\ref{eq:FSRSi0}), the static averaging limit  (\ref{eq:staticlimit}) only contains absorptive line shapes since bath dynamics is neglected during the dephasing time. Furthermore the time evolution in this case is governed by a snapshot of the populations of the excited states.  Eq. (\ref{eq:FSRSi0}) and (\ref{eq:staticlimit}) are therefore expected to be different at short times and become more similar at longer time.

Note that since the signal (\ref{eq:FSRSi0}) is written in terms of Green's functions expanded in sum over states, it can be applied to complex systems with multiple vibrational modes coupled to various baths. In the typical chemical reaction, such as isomerization, only few collective coordinates are involved. Therefore these degrees of freedom are typically treated explicitly whereas the rest of the vibrational and bath degrees of freedom can be treated by a harmonic approximation. In the following we use a simple model system to illustrate the power of SLE approach .
Using Eq. (\ref{eq:FSRSi0})} we performed simulations of the signals for a model system with four vibrational modes and ten bath states ($N=10$). 
Two vibrational states (i.e., $a$ and $c$) are included for each mode, there are two vibrational states (i.e., $a$ and $c$).  We use a kinetic model described by rate equation (\ref{eq:rateeq}) and we assume a linear chain of forward and backward reactions among the bath states.
\begin{equation}
{\rm State\ 1} \overset{k_1}{\underset{k_{-1}}{\rightleftarrows}} {\rm State\ 2} \overset{k_2}{\underset{k_{-2}}{\rightleftarrows}} \cdots \overset{k_9}{\underset{k_{-9}}{\rightleftarrows}} {\rm State\ 10}.
\label{eq:kinetics}
\end{equation}
The rate constants  vary linearly along the chain
\begin{eqnarray}
\begin{cases}
k_i = k_1 + \frac {k_9 - k_1} {8} (i - 1), \\
k_{-i} = 0.1 k_i,
\end{cases}
\label{eq:rateconsts}
\end{eqnarray}
where $k_1>k_9$.  These are given in Table \ref{parameters}.  The process slows down along the chain. This allows to observe both fast and slow jump modulation regimes.  Considering the detailed balance relation, the second line in Eq. (\ref{eq:rateconsts}) indicates a constant energy difference between two neighboring bath states, namely $s$ and $s+1$ states.  The resulting population dynamics, obtained from Eq. (\ref{eq:concentrations}), is depicted in Fig. S1 \cite{supp}. At $T=20$ ps, not only the state 10 but also several bath states contribute to the signal.  As shown in Fig. \ref{frequencychange}, in our model, the frequency for a given vibrational mode depends on the bath states (Eq. (\ref{eq:ls2})), and satisfies a linear relation.
\begin{equation}
\omega_{ca}^{(s)}=\omega_{ca}^{(1)} + \delta_{ca} (s-1),
\label{eq:omegachange} 
\end{equation}
where $\delta_{ca}$ represents the frequency shift when the bath transits from the $s$ to the $s+1$ state.  As shown in Fig. \ref{frequencychange} and Table \ref{parameters}, we employed two parameter regimes both using two values of $\delta_{ca}$; small $\delta_{1}$ is used for two modes (mode 1 and 2, hereafter), and large $\delta_{2}$ is used for the remaining two (mode 3 and 4).  The two regimes correspond to different relation between the jump rate $k$ and splitting $\delta_{ca}$. In the first regime (I), modes 1 and 2 rapidly evolve ( $k_{1} \lesssim \delta_{1}$, fast modulation limit - FML) and the modes 3 and 4 are modulated slowly ($k_{1} \ll \delta_{2}$, SML).  In contrast, in the second parameter regime (II), all four modes are subject to FML.  Note that at longer delay times, the jump rate itself slows down ($k_1>k_9$) as mentioned above.  The $\omega_{ca}^{(1)}$ of mode 2 and 3 are set so that their $\omega_{ca}^{(s)}$ frequencies show crossing between each other  (see Fig. \ref{frequencychange} for the crossing frequencies).  We take $\mathcal{E}_1(t) = \mathcal{E}_1 e^{-i \omega_1 (t-T)}$ to be monochromatic, whereas the Raman probe has a Gaussian envelope with center frequency $\omega_{p}$ and finite duration $\sigma$, $\mathcal{E}_p(t) = \mathcal{E}_p e^{-(t-T)^{2}/2\sigma^{2} -i \omega_p (t-T)}$.  The integrations of Eq. (\ref{eq:FSRSi0}) are then performed analytically to get $S_{FSRS}(\omega,T)$, where prefactor $\sigma^{2} |\mathcal{E}_{1}|^{2} |\mathcal{E}_{a}|^{2} |\mathcal{E}_{p}|^{2} \alpha^{2}_{ac} |V_{ag}|^{2}$ is set to be 1.  Note, that in reality the duration of the $\mathcal{E}_1$ is finite. However since it is in the picosecond range, whereas all dynamical processes as well as the duration of the $\mathcal{E}_p$ are in the femtosecond range, the the CW approximation is justified. In more general case, the bandwidth of $\mathcal{E}_1$ has to be taken into account which will reduce the spectral resolution of the signal. As has been shown in our earlier work \cite{Dorfman:PCCP:2013} the duration of $\mathcal{E}_p$ has to be optimized in order to obtain both high temporal and spectral resolution.

\begin{table}[htbp]
\begin{center}
\begin{threeparttable}
\caption{\label{parameters}Parameters employed in the FSRS simulations.}
\begin{tabular}{lllllll} \toprule
Parameter regime & \phantom{11} $k_{1}$ (s${}^{-1}$) & \phantom{1} $k_{9}$ (s${}^{-1}$) & \phantom{1} $\delta_{1}$ (s${}^{-1}$) \tnote{a} & \phantom{1} $\delta_{2}$ (s${}^{-1}$) \tnote{a} & \phantom{1} $\sigma$ (fs) \tnote{b} & \phantom{1}  $\gamma_{a}$ (s${}^{-1}$) \tnote{c} \\ \midrule
I & \phantom{11} $1.00\times10^{12}$ & \phantom{1} $0.667\times10^{12}$ & \phantom{1} $3.76\times10^{12}$ & \phantom{1} $7.51\times10^{12}$ & \phantom{1} $20.0$ & \phantom{1} $1.88\times10^{12}$ \\
II & \phantom{11} $1.00\times10^{12}$ & \phantom{1} $0.333\times10^{12}$ & \phantom{1} $0.939\times10^{12}$ & \phantom{1} $3.76\times10^{12}$ & \phantom{1} $30.0$ & \phantom{1} $1.88\times10^{12}$ \\ \bottomrule
\end{tabular}
\begin{tablenotes}
\item[a] The frequency shift $\delta_{ca}$ is small $\delta_{1}$ for two normal modes, and large $\delta_{2}$ for the rest two modes.
\item[b] The center frequency of the Gaussian Raman probe, $\omega_{p}$, is equal to $\omega_{1}-1000.0$ cm$^{-1}$.
\item[c]  The vibrational dephasing time corresponds to linewidth of 10.0 cm$^{-1}$, and is used for all four vibrational modes.
\end{tablenotes}
\end{threeparttable}
\end{center}
\end{table}

\begin{figure}[htbp]
\begin{center}
\subfigure[]{\includegraphics[scale=0.6,trim=170 0 100 0]{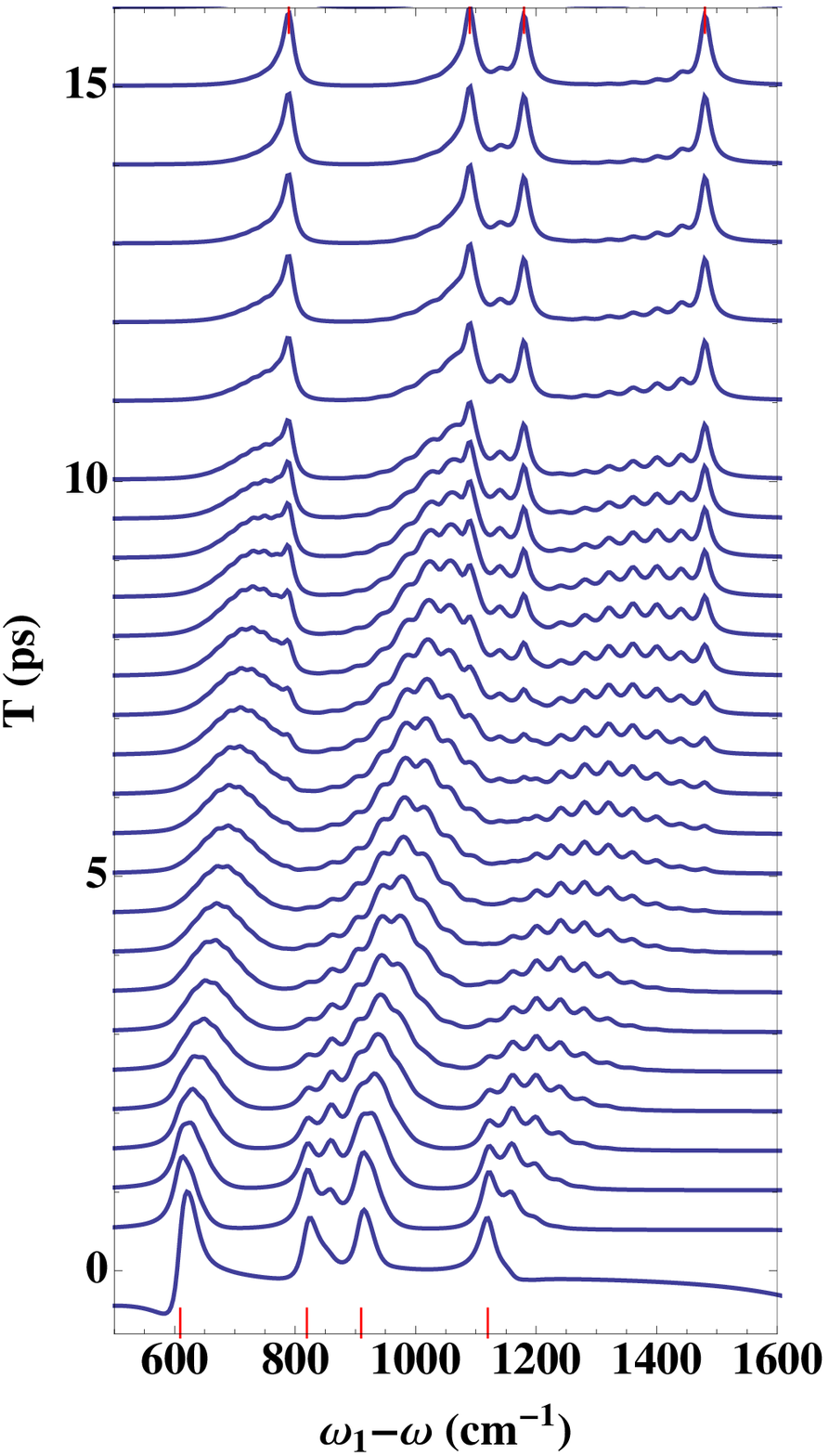}}
\hspace{50mm}
\subfigure[]{\includegraphics[scale=0.642,trim=120 0 100 0]{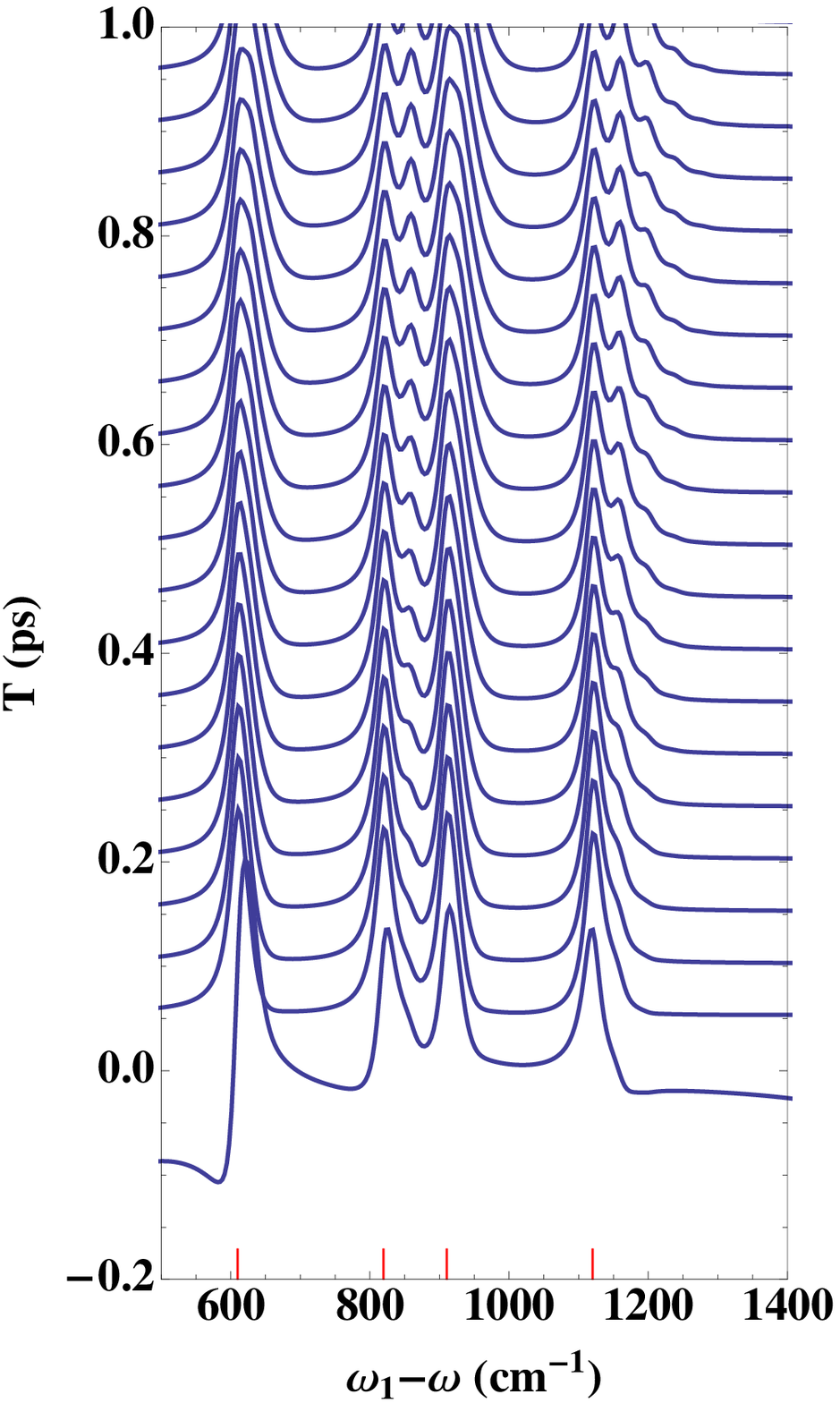}}
\caption{Variation of simulated FSRS signals with different delay times $T$ up to (a) 15 ps and (b) 0.95 ps.  Parameter regime I was employed. After $T=2$ fs, time intervals are (a) 500 fs from $T = 500$ fs up to 10 ps, and 1 ps later and (b) 50 fs from $T = 50$ fs up to 1 ps. 
The stick spectra on the horizontal bottom (top) axis represent the frequencies of the state 1 (state 10).}
\label{FSRS_cond01}
\end{center}
\end{figure}

\begin{figure}[htbp]
\begin{center}
\subfigure[]{\includegraphics[scale=0.6,trim=170 0 100 0]{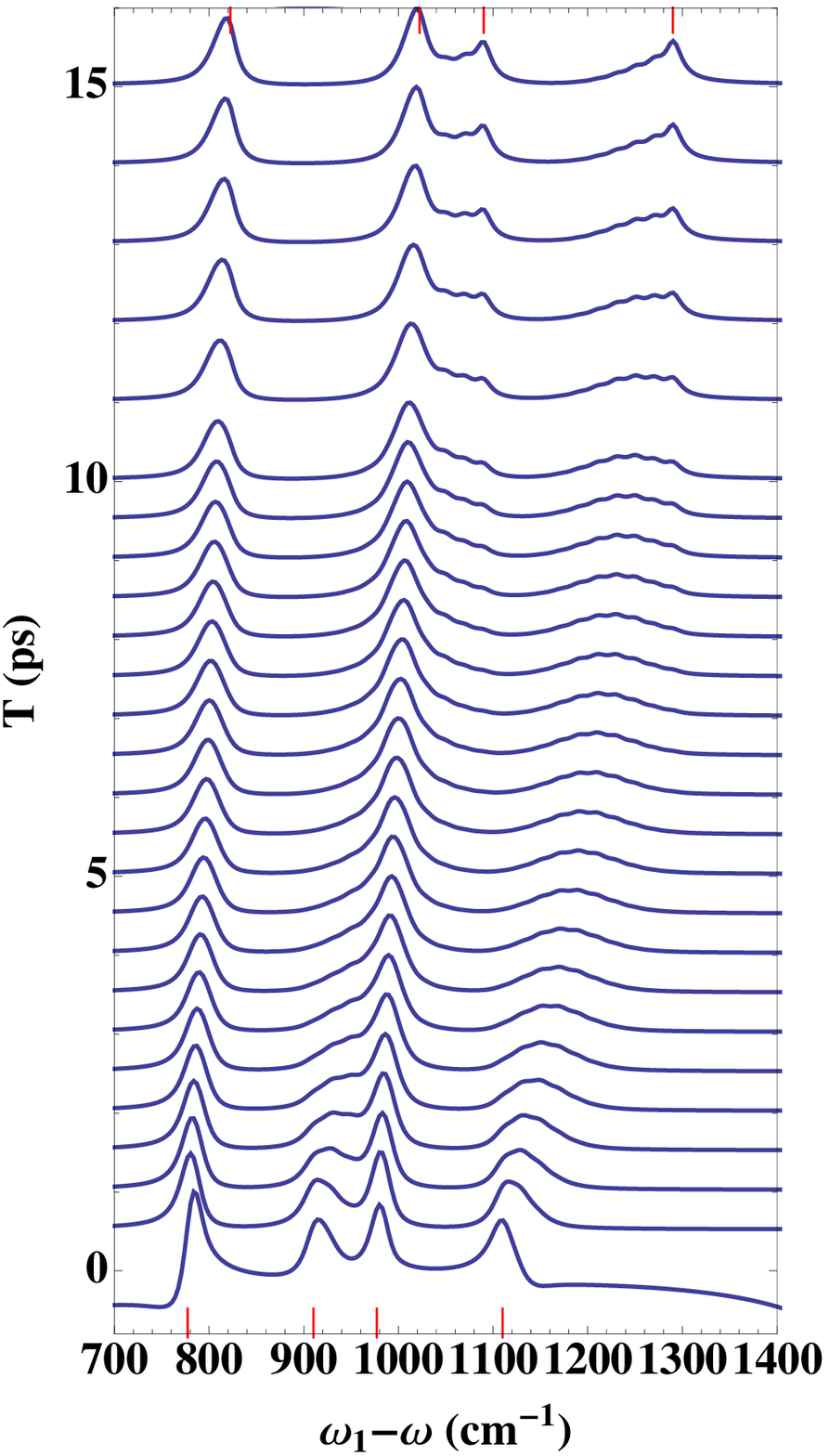}}
\hspace{50mm}
\subfigure[]{\includegraphics[scale=0.642,trim=120 0 100 0]{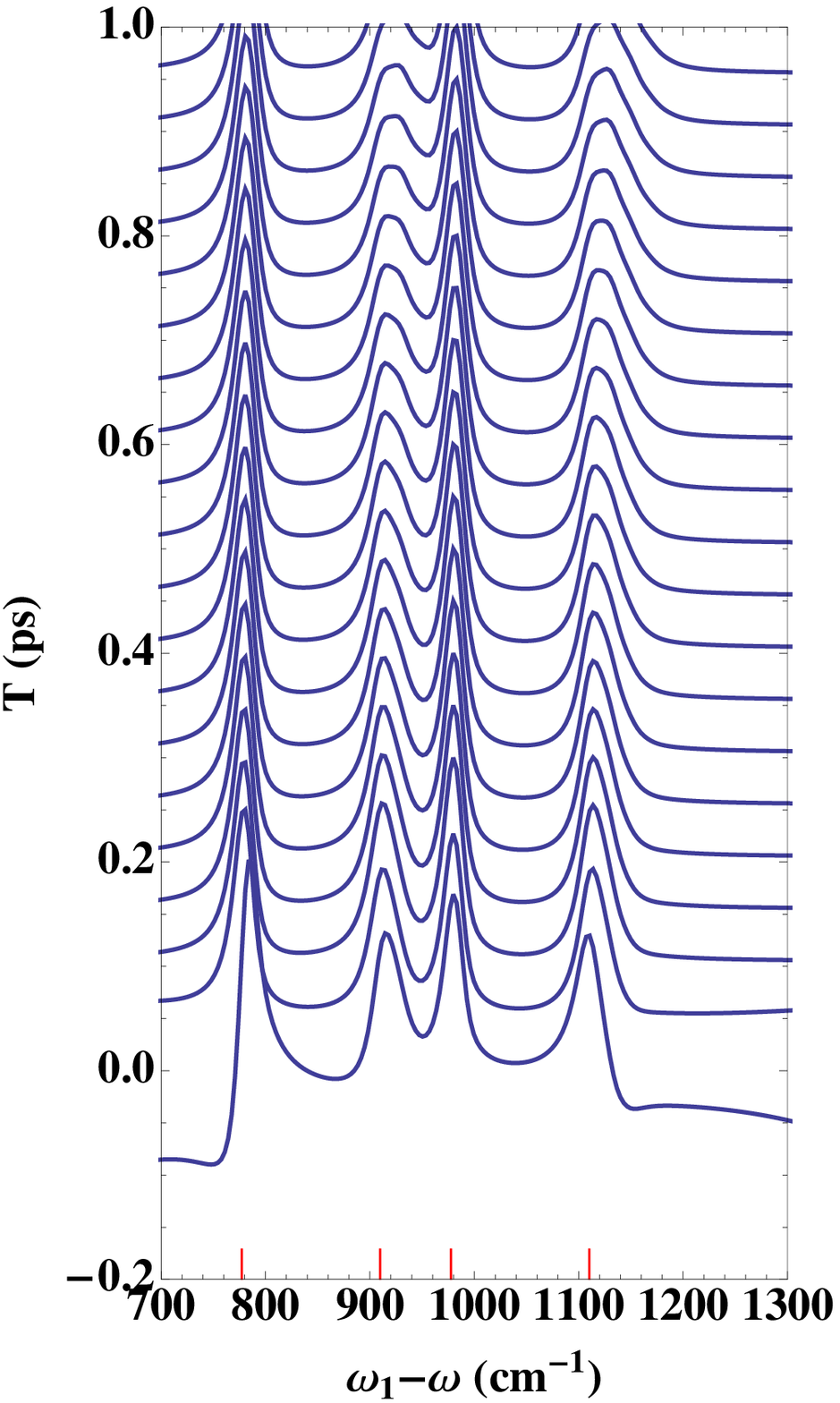}}
\caption{Variation of simulated FSRS signals with different delay times $T$ up to (a) 15 ps and (b) 0.95 ps.  Parameter regime II was employed.  After $T=2$ fs, time intervals are (a) 500 fs from $T = 500$ fs up to 10 ps, and 1 ps later and (b) 50 fs from $T = 50$ fs up to 1 ps.   
The stick spectra on the horizontal bottom (top) axis represent the frequencies of the state 1 (state 10).}
\label{FSRS_cond08}
\end{center}
\end{figure}

The resulting FSRS signals obtained for regime I are shown in Fig. \ref{FSRS_cond01}.  In mode 3 and 4, corresponding to SML, fine structure features of bath states are found even at early delay times (around 1 ps, see Fig. \ref{FSRS_cond01}b).   The fine structure is reduced for longer delay times ($> 10$ ps),  reflecting the population decay shown in Fig. S1 \cite{supp}.  This indicates that the signals are in the snapshot limit. The SLE coincides with the static average limit.  In modes 1 and 2, fine structures are found only after 5 ps.  At $T=2$ fs, dispersive lineshapes are present since the dynamics is not negligible during the dephasing (see Eq. (\ref{eq:staticlimit})).  The FSRS signals for regime II are shown in Fig. \ref{FSRS_cond08}.  At early delay times, the signals have no fine structure (Fig. \ref{FSRS_cond08}b).  Only in mode 3 and 4 at long delay times ($> 5$ ps), there are several weak fine structure features due to approaching the SML.  These signals are beyond the snapshot limit.  Furthermore, as the Raman signal is measured using heterodyne detection which involves terms that correspond to interference between signal and background contributions. Unlike homodyne detection which yields only absorptive line shapes, these interference terms may result in dispersive line shapes. The dispersive line shapes, however do not always show up in the heterodyne detected signals and their existence depend on the parameters of the system as well as laser pulses as we showed above.
Therefore the SLE allows to explain the dispersive line shapes observed when the dynamics of the system is fast compared to the dephasing time scale, that it cannot be neglected. Such simple analysis is only possible in Liouville space and cannot be done in Hilbert space. 

\section{Discussion and Conclusions}
In summary, we have employed a superoperator diagrammatic technique to calculate the linear and third order frequency-dispersed transmission signals. Assuming impulsive (femtosecond) actinic pulse we show that both TASP and FSRS signals can be recast in
terms of a generalized linear response function with respect to pump and probe field
from a non-stationary state prepared by the actinic pulse. Our expressions can be used 
in three types of simulation protocols: Eq.~(\ref{TA-first}) and Eq.~(\ref{FSRS-first}) may be used in direct nonadiabatic propagation of the wavefunctions. The sum 
over states expressions Eq.~(\ref{eq:STA}) and Eq.~(\ref{eq:FSRS2}) may be used for the interpretation of resonances. Finally the stochastic Liouville equations (Eq.~\ref{eq:FSRSi0}-\ref{eq:staticlimit}) provide  an inexpensive level of modeling that is numerically less demanding than the direct propagation. This approach provides a qualitative explanation for the dispersive features observed in Raman signals. It further covers the fast dynamical behavior which cannot be described the snapshot limit, which yields the system response far beyond the instantaneous frequency tracking.


 We first considered the linear analogue of the FSRS - a transient absorption of the shaped (broad plus narrowband) pulse  $S_{TASP}(\omega,T)$ signal: linear transmission of a single probe following a femtosecond actinic-pump pulse. The entire process (including the actinic pulse) is a $\chi^{(3)}$ which can be interpreted as a generalized $\tilde{\chi}^{(1)}$ when starting in a nonstationary state. The signal with a second order (three wave mixing) probe is a $\chi^{(4)}$ process which can be recast as a generalized $\tilde{\chi}^{(2)}$. For a third oder - four wave mixing (FWM) probe the  $\chi^{(5)}$ process can be described as a generalized $\tilde{\chi}^{(3)}$. The generalized  $n$-th order susceptibilities $\tilde{\chi}^{(n)}$ depend on $n+1$  rather than  $n$ independent frequency  variables due to the lack of time translational invariance of the time evolution of the system. 

The present formalism can be generalized to other  spectroscopic  techniques. The off-resonant FSRS can be extended to the resonant case, which generally provides a stronger and more complex signal. In particular the six-wave mixing process which generally described by $\chi^{(5)}$ will be reduced to a four-wave mixing described by a generalized $\tilde{\chi}^{(3)}$ compared to a generalized $\tilde{\chi}^{(1)}$ of a combined broad plus narrowband pulse in the off-resonant case. The
Impulsive excitation may include  photoexcitation by a short optical or infrared  pulse \cite{ Proceeding,Fayer,Shaulbook}, photoionization \cite{photoelectron} that prepares an excited ion and core excitation followed by an Auger process \cite{XPS}  that leaves the molecule in a doubly ionized state. 

\section*{Acknowledgements}
We gratefully acknowledge the support of the National Science
Foundation (NSF) through grant no. CHE-1361516, (U.S.) Department of Energy (DOE) and the
National Institute of Health (NIH) grant no. GM-59230.

\section*{Appendix A: Derivation of the TASP signal}
We read the signal from  the diagrams of Fig.~1. First we expand the exponential in Eq.~(\ref{first-definition}) to the first order ${\cal
E}_1$ and ${\cal E}_p$ and second order in ${\cal E}_a$ which yield for the first diagram (i)
\bea
S_{TASP}^{(i)}(\omega; t_0,\tau_0)&\!=\!&\! \frac{2}{\hbar} \,{\rm Im} \Big[ \int_{-\infty}^{\infty} dt
\int_{-\infty}^{t} d\tau_3 \int_{-\infty}^{\tau_3} d\tau_2 \int_{-\infty}^{\tau_2} d\tau_1
\,e^{i \omega (t-t_0)} \,{\cal E}_{p}^*(\omega) \,{\cal E}_{1}(\tau_3\!-\!t_0) \nonumber \\
&&\Big[\langle V_L {\cal G}(t\!-\!\tau_3) V_{R}^{\dagger} {\cal G}(\tau_3\!-\!\tau_2) V_{R} {\cal
G}(\tau_2\!-\!\tau_1) V_{L}^{\dagger} \rangle \,{\cal E}_{a}^{*} (\tau_2\!-\!\tau_0) \,{\cal
E}_{a}(\tau_1\!-\!\tau_0) \nonumber 
\\&+& \langle V_L {\cal G}(t\!-\!\tau_3) V_{R}^{\dagger} {\cal G}(\tau_3\!-\!\tau_2) V_{L}^{\dagger} {\cal
G}(\tau_2\!-\!\tau_1) V_{R} \rangle \,{\cal E}_{a}(\tau_2\!-\!\tau_0)\, {\cal
E}_{a}^{*}(\tau_1\!-\!\tau_0)\Big].
\eea 
where two terms indicate two possible ways that actinic pulse can excite the ground state to the vibrational state of excited electronic state. For the first (second) term the absorption happens first on the ket (bra) side followed by the absorption on the bra (ket) side.
Now assuming the actinic pulses to be impulsive ${\cal E}_a(\tau)= {\cal E}_a
\delta(\tau)$, we get 
\bea
S_{TASP}^{(i)}(\omega;t_0,\tau_0)&\!=\!& \!\frac{2}{\hbar} \,{\rm Im} \Big[
(-\frac{i}{2\hbar})\int_{-\infty}^{\infty}
dt \int_{0}^{\infty} dt_3 \,e^{i \omega (t-t_0)}
\,\Big[\langle V_L {\cal G}(t_3) V_{R}^{\dagger} {\cal G}(t\!-\!t_3\!-\!\tau_0) V_{R}  V_{L}^{\dagger} \rangle   \nonumber \\
&& + \langle V_L {\cal G}(t_3) V_{R}^{\dagger} {\cal G}(t\!-\!t_3\!-\!\tau_0) V_{L}  V_{R}^{\dagger} \rangle \Big] \,{\cal E}_{p}^{*}(\omega) {\cal E}_{1}(t\!-\!t_3\!-\!t_0) |{\cal E}_{a}|^{2}\Big],
\eea
where note that the first propagator ${\cal G}$ from the right is evaluated at $t=0$ and
we use  ${\cal G}(0)=-i /2\hbar$.  The action of $V_R V_L^{\dagger}$ or $V_L^{\dagger} V_R$ on ground electronic
state creates vibrational wave-packet in the excited state $|a c\rangle \rangle$ which is a non-stationary state. Now
performing inverse Fourier transformation for the propagator and for the electric field
and integrating over $t$ and $t_1$ variables we obtain the signal as 
\bea
S_{TASP}^{(i)}(\omega;t_0,\tau_0)&=& \frac{2}{\hbar} \,{\rm Im} \Big[(-\frac{i}{\hbar})
\int_{-\infty}^{\infty}
\frac{d\omega_0}{2 \pi} \int_{-\infty}^{\infty} \frac{d\omega_1'}{2\pi} \big \langle V_L
{\cal
G}(\omega_1'+\omega_0) V_{R}^{\dagger}{\cal G}(\omega_0))\big \rangle^{'}
e^{-i(\omega-\omega_1')t_0} e^{i \omega_0 \tau_0}
\nonumber \\
&&\times\,
{\cal E}_{p}^{*}(\omega) {\cal E}_{1}(\omega_1') |{\cal E}_a|^2 2\pi \delta(\omega-\omega_0-\omega_1')\Big] .
\eea
Finally integrating out $\omega_0$ variable we obtain 
\be 
S_{TASP}^{(i)}(\omega;t_0-\tau_0)= \frac{2}{\hbar} \, {\rm Im}
\Big[(-\frac{i}{\hbar})
\int_{-\infty}^{\infty}
\frac{d\omega_1'}{2\pi} \big \langle V_L {\cal
G}(\omega) V_{R}^{\dagger}{\cal G}(\omega-\omega_1')\big \rangle\rangle^{'} {\cal
E}_{p}^{*}(\omega) {\cal E}_{1}(\omega_1') |{\cal E}_a|^2 
e^{-i(\omega-\omega_1')(t_0-\tau_0)}\Big].
\ee
Note that the above expression depends on the time delay between the actinic and the
probe pulses which we write as $T=t_0-\tau_0$. Therefore the signal can be recast as 
\be
S_{TASP}^{(i)}(\omega;T)= \frac{2}{\hbar} \, {\rm Im} \Big[(-\frac{i}{\hbar})
\int_{-\infty}^{\infty}
\frac{d\Delta}{2\pi} \,
e^{-i\Delta T} {\cal
E}_{p}^{*}(\omega) {\cal E}_{1}(\omega+\Delta)
\tilde{\chi}_{TASP(i)}^{(1)}(-\omega,\omega+\Delta)\Big],
\label{final-linear-signal}
\ee
where the generalized susceptibility $\tilde{\chi}_{TASP(i)}^{(1)}(-\omega,\omega+\Delta)$ is given
as 
\be
\tilde{\chi}_{TASP(i)}^{(1)}(-\omega,\omega+\Delta) = 
\langle V_L {\cal
G}(\omega) V_{R}^{\dagger}{\cal
G}(-\Delta)\rangle^{'} .
\label{chi1}
\ee
The expression for diagram (ii) of Fig.~1 can be derived similarly.

\section*{Appendix B: Three and Four-Wave-Mixing Probes}
Here we consider  a three wave-mixing (TWM) and four wave-mixing (FWM) signal. After preparation of the state $\rho_{ac}$ we sent pulses $\mathcal{E}_1$, $\mathcal{E}_2$ for TWM, and additional $\mathcal{E}_3$ for FWM which are centered around $T_1$, $T_2$, and $T_3$ relative to the preparation time $\tau_0$. We then detect the frequency dispersed transmission of the probe pulse $\mathcal{E}_p$ centered at $t=T$ with respect to $\tau_0$. The signal can be calculated in the second and third order of the field matter interactions for TWM and FWM respectively and we obtain 
\begin{align}
S_{TWM}(\omega;T,T_1,T_2)= \frac{2}{\hbar} \, {\rm Im} \Big[&
\int_{-\infty}^{\infty}
\frac{d\omega_1'}{2\pi} \int_{-\infty}^{\infty} \frac{d\omega_2'}{2\pi}\,
e^{-i\omega T+i\omega_1'T_1+i\omega_2' T_2}\notag\\
&\times {\cal
E}_{p}^{*}(\omega) \tilde{{\cal E}}_{1}(\omega_1')\tilde{{\cal E}}_{2}(\omega_2')
\tilde{\chi}_{TWM}^{(2)}(-\omega,\omega_1',\omega_2')\Big],
\label{eq:TWM}
\end{align}
where the second order generalized susceptibility is given by 
\be
\tilde{\chi}_{TWM}^{(2)}(-\omega,\omega_1',\omega_2') = \langle \tilde{V}_L {\cal
G}(\omega) \tilde{V}_{-}{\cal
G}(\omega -\omega_2')\tilde{V}_{-} {\cal G}(\omega-\omega_2'-\omega_1') \rangle^{'}.
\label{chi2}
\ee
Similarly the FWM signal is given by 
\begin{align}
S_{FWM}(\omega;T,T_1,T_2,T_3)= \frac{2}{\hbar} \, {\rm Im} \Big[& \int_{-\infty}^{\infty}
\frac{d\omega_1'}{2\pi}\, \int_{-\infty}^{\infty} \frac{d\omega_2'}{2\pi} \int_{-\infty}^{\infty} \frac{d\omega_3'}{2\pi} \,
e^{-i\omega T+i\omega_1'T_1+i\omega_2'T_2+i\omega_3'T_3}\notag\\
&\times {\cal
E}_{p}^{*}(\omega) \tilde{{\cal E}}_{1}(\omega_1')\tilde{{\cal E}}_{2}(\omega_2') {\tilde{\cal E}}_{3}(\omega_3')
\tilde{\chi}_{FWM}^{(3)}(-\omega,\omega_1',\omega_2',\omega_3')\Big],
\label{eq:FWM}
\end{align}
where the third order generalized susceptibility is 
\be
\tilde{\chi}_{FWM}^{(3)}(-\omega,\omega_1',\omega_2',\omega_3') =\langle \tilde{V}_L {\cal
G}(\omega) \tilde{V}_{-}{\cal
G}(\omega-\omega_3')\tilde{V}_-\mathcal{G}(\omega-\omega_3'-\omega_2')\tilde{V}_- {\cal G}(\omega-\omega_3'-\omega_2'-\omega_1')\rangle^{'}, 
\label{chi3}
\ee
where $\tilde{V}=V+V^{\dagger}$ is the full matter transition operator. Note, that Eq. (\ref{eq:TWM}) and Eq.~(\ref{eq:FWM}) do not rely on the rotating wave approximation (RWA).


\begin{thebibliography}{99}
\bibitem{Miz97} Y, Mizutani and T. Kitagawa, Science {\bf 278}, 443 (1997).
\bibitem{McCamant:JPCA:2003} D. W. McCamant, P. Kukura, and R. A. Mathies, J. Phys. Chem. A {\bf 107}, 8208 (2003).
\bibitem{Lee:JCP:2004} S.-Y. Lee, D. Zhang, D. W. McCamant, P. Kukura, and R. A. Mathies, J. Chem. Phys {\bf 121}, 3632 (2004).
\bibitem{Fang:Nature:2009} C. Fang, R. R. Frontiera, R. Tran, and R. A. Mathies, Nature {\bf 462}, 200 (2009).
\bibitem{Kukura:Science:2005} P. Kukura, D. W. McCamant, S. Yoon, D. B. Wandschneider and R. A. Mathies, Science,  {\bf 310}, 1006 (2005).
\bibitem{Kukura:AnnurevPhysChem:2007} P. Kukura, D. W. McCamant and R. A. Mathies, Annu. Rev. Phys. Chem., {\bf 58}, 461 (2007).
\bibitem{Takeuchi:Science:2008} S. Takeuchi, S. Ruhman, T. Tsuneda, M. Chiba, T. Taketsugu, and T. Tahara, Science {\bf 322}, 1073 (2008).
\bibitem{Kur12} H. Kuramochi, S. Takeuchi, and T. Tahara, J. Phys. Chem. Lett. {\bf 3}, 2025 (2012).
\bibitem{Zan09}S.-H. Shim and M. T. Zanni, Phys. Chem. Chem. Phys. {\bf 11}, 748 (2009).
\bibitem{biggs12} J. D. Biggs, Y. Zhang, D. Healion, and S. Mukamel, J. Chem. Phys {\bf 136}, 174117 (2012).
\bibitem{biggs13} S. Mukamel, D. Healion, Y. Zhang, and J. D. Biggs, Annu. Rev. Phys. Chem. {\bf 64}, 101 (2013).
\bibitem{Mohammed:Science:2005} O. F. Mohammed, D. Pines, J. Dreyer, E. Pines, and E. T. J. Nibbering, Science {\bf 310}, 83 (2005). 
\bibitem{Schreier:Science:2007} W. J. Schreier, T. E. Schrader, F. O. Koller, P. Glich, C. E. Crespo-Hernandez, V. N. Swaminathan, T. Carell, W. Zinth, and B. Kohler Science {\bf 315} 625 (2007).
\bibitem{Liebel:arXiv:2013} M. Liebel, C. Schnedermann, and P. Kukura, Phys. Rev. Lett {\bf 112}, 198302 (2014).
\bibitem{McCamant:RevSci:2004} D. W. McCamant, P. Kukura, S. Yoon, and R. A. Mathies, Rev. Sci. Instr. {\bf 75}, 4971 (2004).
\bibitem{Kraack:PCCP:2013} J. P. Kraack, A. Wand, T. Buckup, M. Motzkus, and S. Ruhman, Phys. Chem. Chem. Phys. {\bf 15}, 14487 (2013).
\bibitem{Yoshizawa: 1999} M. Yoshizawa, M. Kurosawa, Phys. Rev. A, {\bf 61}, 013808 (1999).
\bibitem{Sun:2008} Z. Sun, J. Lu, D. H. Zhang, S.-Y. Lee, J. Chem. Phys, {\bf 128}, 144114 (2008).
\bibitem{Zhao:2011} B. Zhao, Z. Sun, S. -Y. Lee, J. Chem. Phys, {\bf 134}, 024307 (2011).
\bibitem{Dorfman:PCCP:2013} K.E. Dorfman, B.P. Fingerhut, and S. Mukamel, Phys. Chem. Chem. Phys, {\bf 15}, 12348, (2013).
\bibitem{Kubo1} R. Kubo, {\it Fluctuations, Relaxation and Resonance in Magnetic Systems}
(Oliver and Boyd, Edinburgh, 1962), p. 23.
\bibitem{Kubo2} R. Kubo, J. Math. Phys. {\bf 4}, 174 (1963).
\bibitem{And54} P. W. Anderson, J. Phys. Soc. Jpn. {\bf 9} 316 (1954).
\bibitem{Dorfman:JCP:2013} K.E. Dorfman, B.P. Fingerhut, and S. Mukamel J. Chem. Phys.
{\bf 139}, 124113 (2013).
\bibitem{Ando:2014} H. Ando, B. P. Fingerhut, K. E. Dorfman, J. D. Biggs, S. Mukamel, J. Am. Chem. Soc, {\bf 136} 14801 (2014). 
\bibitem{Harbola} C. Marx, U. Harbola, S. Mukamel, Phys. Rev. A  {\bf 77}, 022110
(2008).
\bibitem{Shaulbook} S. Mukamel, {\it Principles of Nonlinear Optics and Spectroscopy},
Oxford University Press, Oxford, UK, (1995).
\bibitem{Rahav_review} S. Rahav and S. Mukamel, Adv. At., Mol., Opt. Phys.,
{\bf 59}, 223 (2010).

\bibitem{Kumar:2001} A.T.N Kumar, F. Rosca, A. Widom, P.M. Champion J. Chem. Phys. {\bf 114}, 6795 (2001).
\bibitem{Kumar2:2001} A.T.N Kumar, F. Rosca, A. Widom, P.M. Champion J. Chem. Phys. {\bf 114}, 701 (2001).

\bibitem{Dantus1} V. V. Lozovoy, I. Pastirk, K. A. Walowicz, and M. Dantus, J. Chem. Phys. {\bf 118}, 3187-3196 (2003).
\bibitem{Dantus2}V. V. Lozovoy, I. Pastirk, and M. Dantus, Opt. Lett. {\bf 29}, 775-777 (2004).
\bibitem{Dantus3} A. Konar, J. Shah, V. V. Lozovoy and M. Dantus, J. Phys. Chem. Lett.  {\bf 3}, 1329-1335 (2012).
\bibitem{Dantus4} A. Konar, V. V. Lozovoy and M. Dantus, J. Phys. Chem. Lett. {\bf 3}, 2458-2464 (2012).
\bibitem{Shaul_control} S. Mukamel, J. Chem. Phys. {\bf 139}, 164113 (2013).
\bibitem{photo2} K. Adamczyk, M. Premont-Schwarz, D. Pines, E. Pines and
E. T. J. Nibbering, Science,  {\bf 326}, 1690-1694 (2009).
\bibitem{Marek}  M. S. Marek, T. Buckup, J. Southall, R. J. Cogdell and M. Motzkus, J. Chem. Phys. {\bf 139},
074202 (2013).
\bibitem{Kraack} J. P. Kraack,  T. Buckup and M. Motzkus, Phys. Chem. Chem. Phys., {\bf 14}, 13979-13988 (2012).

\bibitem{Sanda} F. Sanda and S. Mukamel, J. Phys. Chem. B {\bf 112}, 14212 (2008).
\bibitem{Gamliel} D. Gamliel and H. Levanon, {\it Stochastic Processes in Magnetic Resonance}
(World Scientific Publishing Company, Inc., 1995).
\bibitem{Tanimura} Y. Tanimura, J. Phys. Soc. Jpn. {\bf 75}, 082001 (2006).
\bibitem{supp} See supplementary material for the plots of kinetic evolution of the bath states.
\bibitem{Proceeding} Proceedings of the 19th conference on ultrafast phenomena,  K Yamanouchi, S Cundiff,  M K Gonokami,  L DiMauro, Springer (2015).
\bibitem{Fayer} M. D. Fayer, {\it Ultrafast Infrared Vibrational Spectroscopy}, CRC
Press, Taylor and Francis Group, USA, 2013.

\bibitem{photoelectron} A. L. Thompson and T. J. Martinez, Faraday Discuss. {\bf 150}, 293 (2011).
\bibitem{XPS} B. K. McFarland et al. Nat. Comm., {\bf 5}, 4235 (2014).




  
\end{thebibliography}
\end{document}